\documentclass[twocolumn]{aastex631}
\UseRawInputEncoding
\usepackage[utf8]{inputenc}
\usepackage[T1]{fontenc}

\usepackage{hhline}

\usepackage{booktabs}
\usepackage{footmisc}
\usepackage{threeparttable}
\usepackage{amsmath}
\usepackage{tabularx}
\usepackage{hyperref}

\begin{document}
\title{Applying multimodal learning to Classify transient Detections Early \textnormal{\texttt{(AppleCiDEr)}} I: Data set, methods, and infrastructure}

\correspondingauthor{Alexandra Junell}
\email{ajunell@umn.edu}

\correspondingauthor{Argyro Sasli}
\email{asasli@umn.edu}

\author[0000-0002-9380-7983]{Alexandra Junell}
\affiliation{School of Physics and Astronomy, University of Minnesota, Minneapolis, MN 55455, USA}
\affiliation{NSF Institute on Accelerated AI Algorithms for Data-Driven Discovery (A3D3)}

\author[0000-0001-7357-0889]{Argyro Sasli}
\affiliation{School of Physics and Astronomy, University of Minnesota, Minneapolis, MN 55455, USA}
\affiliation{NSF Institute on Accelerated AI Algorithms for Data-Driven Discovery (A3D3)}

\author[0000-0001-7129-1325]{Felipe Fontinele Nunes}
\affiliation{School of Physics and Astronomy, University of Minnesota, Minneapolis, MN 55455, USA}
\affiliation{NSF Institute on Accelerated AI Algorithms for Data-Driven Discovery (A3D3)}

\author{Maojie Xu}
\affiliation{Department of Computer Science \& Engineering, University of Minnesota, Minneapolis, MN 55455, USA}
\affiliation{NSF Institute on Accelerated AI Algorithms for Data-Driven Discovery (A3D3)}

\author{Benny Border}
\affiliation{School of Physics and Astronomy, University of Minnesota, Minneapolis, MN 55455, USA}
\affiliation{NSF Institute on Accelerated AI Algorithms for Data-Driven Discovery (A3D3)}

\author[0000-0002-5683-2389]{Nabeel~Rehemtulla}
\affiliation{Department of Physics and Astronomy, Northwestern University, 2145 Sheridan Road, Evanston, IL 60208, USA}
\affiliation{Center for Interdisciplinary Exploration and Research in Astrophysics (CIERA), 1800 Sherman Ave., Evanston, IL 60201, USA}
\affiliation{NSF-Simons AI Institute for the Sky (SkAI), 172 E. Chestnut St., Chicago, IL 60611, USA}

\author{Mariia Rizhko}
\affiliation{University of California, Berkeley, Department of Astronomy, Berkeley, CA, USA}

\author[0000-0003-3658-6026]{Yu-Jing Qin}
\affiliation{Division of Physics, Mathematics and Astronomy, California Institute of Technology, 1200 E California Blvd., Pasadena, CA 91125, USA}

\author[0009-0003-6181-4526]{Theophile Jegou Du Laz}
\affiliation{Division of Physics, Mathematics, and Astronomy, California Institute of Technology, Pasadena, CA 91125, USA}
\affiliation{NSF Institute on Accelerated AI Algorithms for Data-Driven Discovery (A3D3)}

\author{Antoine Le Calloch}
\affiliation{School of Physics and Astronomy, University of Minnesota, Minneapolis, MN 55455, USA}
\affiliation{NSF Institute on Accelerated AI Algorithms for Data-Driven Discovery (A3D3)}

\author[0000-0003-1314-4241]{Sushant Sharma Chaudhary}
\affiliation{School of Physics and Astronomy, University of Minnesota, Minneapolis, MN 55455, USA}
\affiliation{NSF Institute on Accelerated AI Algorithms for Data-Driven Discovery (A3D3)}

\author[0000-0002-1163-2834]{Shaowei Wu}
\affiliation{School of Physics and Astronomy, University of Minnesota, Minneapolis, MN 55455, USA}

\author[0000-0003-1546-6615]{Jesper Sollerman}
\affiliation{Department of Astronomy, Stockholm University, 10691 Stockholm, Sweden}

\author{Niharika Sravan}
\affiliation{Department of Physics, Drexel University, Philadelphia, PA 19104, USA}

\author[0000-0001-5668-3507]{Steven L. Groom}
\affiliation{IPAC, California Institute of Technology, 1200 E. California
             Blvd, Pasadena, CA 91125, USA}

\author{David Hale}
\affiliation{Caltech Optical Observatories, California Institute of Technology, Pasadena, CA  91125}

\author[0000-0002-5619-4938]{Mansi M. Kasliwal}
\affiliation{California Institute of Technology, 1200 E. California Boulevard, Pasadena, CA 91125, USA}

\author[0000-0003-1227-3738]{Josiah Purdum}
\affiliation{Cahill Center for Astrophysics, California Institute of Technology, Pasadena, CA, 91125, USA}

\author[0000-0002-9998-6732]{Avery Wold}
\affiliation{IPAC, California Institute of Technology, 1200 E. California Blvd, Pasadena, CA 91125, USA}

\author[0000-0002-3168-0139]{Matthew J. Graham}
\affiliation{Cahill Center for Astrophysics, California Institute of Technology, Pasadena, CA, 91125, USA}
\affiliation{NSF Institute on Accelerated AI Algorithms for Data-Driven Discovery (A3D3)}

\author[0000-0002-8262-2924]{Michael W. Coughlin}
\affiliation{School of Physics and Astronomy, University of Minnesota, Minneapolis, MN 55455, USA}
\affiliation{NSF Institute on Accelerated AI Algorithms for Data-Driven Discovery (A3D3)}

\begin{abstract}

Modern time-domain surveys like the Zwicky Transient Facility (ZTF) and the Legacy Survey of Space and Time (LSST) generate hundreds of thousands to millions of alerts, demanding automatic, unified classification of transients and variable stars for efficient follow-up. We present \texttt{AppleCiDEr} (Applying Multimodal Learning to Classify Transient Detections Early), a novel framework that integrates four key data modalities (photometry, image cutouts, metadata, and spectra) to overcome limitations of single-modality classification approaches. Our architecture introduces \textit{(i)} two transformer encoders for photometry, \textit{(ii)} a multimodal convolutional neural network (CNN) with domain-specialized metadata towers and Mixture-of-Experts fusion for combining metadata and images, and \textit{(iii)} a CNN for spectra classification. Training on $\sim$ 30,000 real ZTF alerts, \texttt{AppleCiDEr} achieves high accuracy, allowing early identification and suggesting follow-up for rare transient spectra. The system provides the first unified framework for both transient and variable star classification using real observational data, with seamless integration into brokering pipelines, demonstrating readiness for the LSST era.

\end{abstract}


\section{Introduction}
Large-scale surveys in modern astronomy prioritize automated detection \citep{Rebbapragada_2008}, which demands innovative approaches to handle the unprecedented volume and variety of astronomical data \citep{Zhang-2015, app12126202}. A wide range of multiwavelength facilities -- such as the Neil Gerhels Swift Observatory \citep[Swift;][]{Gehrels_2004}, the Palomar Transient Factory \citep{Law_2009}, Advanced LIGO \citep{Aasi_2015}, Catalina Real-Time Transient Survey \citep{2012IAUS..285..306D}, Asteroid Terrestrial-impact Last Alert System \citep{2018PASP..130f4505T}, the Zwicky Transient Facility \citep[ZTF;][]{2019PASP..131a8002B, 2019PASP..131g8001G,2019PASP..131a8003M, 2020PASP..132c8001D}, Einstein Probe \citep{Yuan_2022}, Nancy Grace Roman Space Telescope \citep{10.1117/1.JATIS.6.4.046001}, Ultraviolet Transient Astronomy Satellite \citep{Shvartzvald_2024}, Vera C. Rubin Observatory's Legacy Survey of Space and Time \citep[LSST;][]{2019ApJ...873..111I}, Ultraviolet Explorer \citep{kulkarni2023scienceultravioletexploreruvex}, and future Laser Interferometer Space Antenna \citep{Amaro_Seoane_2023} -- offers opportunities to detect events across multiple detectors and wavelengths. However, enabling Multi-Messenger Astronomy (MMA) depends on robust software ecosystems and fast alerts.

Critical infrastructure components include alert brokers (e.g., \texttt{Fink}; \citealp{10.1093/mnras/staa3602}, \texttt{ALeRCE}; \citealp{Forster2021}, and \texttt{ANTARES}; \citealp{Matheson_2021}), among others, for data processing and distribution, as well as Target Observation Managers (TOMs) such as \texttt{SkyPortal} \citep{vanderWalt2019, 2023ApJS..267...31C} that coordinate follow-up resources. Early classification is paramount for optimizing spectroscopic follow-up of rare transients such as Tidal Disruption Events (TDEs), kilonovae, and Fast Blue Optical Transients (FBOTs). An example of this is the use of available spectra to prevent misclassification, as demonstrated by \cite{gonz_spectra}, which identified 34 misclassified supernovae and revised the estimated core-collapse supernova rate.

In MMA contexts, rapid classification significantly reduces candidate pools for gravitational-wave follow-up \citep{10.1093/mnras/staa1776}. Although fast transients can be discovered without external triggers (e.g. ZTFReST;  \citealp{Andreoni_2021}), current limitations are evident. ZTF's Survey I (2018–2020) secured spectroscopy for 30 TDEs \citep{2023ApJ...942....9H}, whereas FBOTs are often found in archival searches where follow-up is impractical \citep{Ho_2023}.

ZTF already generates $\sim 10^5 - 10^6$ alerts each night \citep{2019PASP..131a8001P}, and LSST \citep{2019ApJ...873..111I} is poised to scale this to $\sim 2 \times 10^7$. Therefore, robust machine learning (ML) frameworks are essential for timely classification, anomaly detection, and deeper archival data exploration. 

Existing classification models have historically specialized along source-type boundaries. Some focus on periodic or quasi-periodic variable stars (for example, see \citealt{Jamal_2020, rizhko2024astrom, Chaini_2024, zuo2025falcofoundationmodelastronomical}), while others target explosive transients such as supernovae (for example, see \citealt{Villar2019, Hosseinzadeh_2020, Fremling2021SNIascore, Pimentel2023, Zhang_2024, sharma2025ccsnscoremultiinputdeeplearning}). Very few frameworks aim to classify both regimes simultaneously; a major limitation for real-time pipelines processing heterogeneous alert streams.

Some efforts, such as the Real-time Automated Photometric IDentification \citep[\texttt{RAPID};][]{2019PASP..131k8002M} and follow-up
work \citep{10.1093/rasti/rzae054}, have attempted mixed-source classification, but rely exclusively on simulated ZTF light curves. LSST-driven initiatives like the Photometric LSST Astronomical Time-series Classification Challenge \citep[\texttt{PLAsTiCC};][]{PLAsTiCC2018, 2019PASP..131i4501K, Malz_2019, Hlozek_2023} and its successor \texttt{ELAsTiCC} pushed the field forward, fostering the development of transformer-based classifiers \citep{10.1093/rasti/rzad046, CabreraVives2024} and highlighting the importance of anomaly detection for rare source discovery \citep{Villar_2021}. However, these benchmark datasets were also based entirely on simulation. 

Practical classifiers integrated into broker pipelines (e.g., \texttt{Fink}; \citealp{10.1093/mnras/staa3602}, \texttt{ALeRCE}; \citealp{S_nchez_S_ez_2021}, and \texttt{ANTARES}; \citealp{Narayan_2018,Matheson_2021}) have been optimized for real-time use, but are typically restricted to single-modality inputs, most often photometric light curves. There are already some multi-modal classifiers in production, for example \cite{10650191, Rehemtulla+2025, duev2021phenomenologicalclassificationzwickytransient, CarrascoDavis2021}. However, the classification focus differs from that in this paper. For example, in \cite{10650191} the primary interest is in distinguishing moving objects and filtering spurious detections rather than a comprehensive astrophysical typing.

In the multimodal domain, models like \texttt{AstroCLIP} (for galaxy images and spectra; \citealp{Parker_2024}), \texttt{PAPERCLIP} (for image–text matching; \citealp{mishrasharma2024paperclipassociatingastronomicalobservations}), and \texttt{LAVA} (for unifying multimodal astronomical data streams; \citealp{zaman2025astrollavaunificationastronomicaldata} represent a shift toward more generalized frameworks. Stellar classification models \citep{10.1093/mnras/stad3015} have similarly adopted CLIP-style contrastive training \citep{radford2021learningtransferablevisualmodels}.

More recent advances in foundation models offer new opportunities for generalization and transfer. Time-domain models like \texttt{AstroM}$^3$ \citep{rizhko2024astrom} and \texttt{MAVEN} \citep{Zhang_2024} demonstrate strong performance within specific source domains, but none currently support a unified treatment of both transients and variable stars using real observational data (i.e. that these do not demonstrate cross-survey functionality). Moreover, they do not incorporate all four key input modalities (photometry, images, metadata, and spectra) which are critical for early-time decision making. This fragmentation limits the utility of current systems in real-time alert pipelines and impedes retrospective discovery, as highlighted by recent archival studies uncovering missed transients in ZTF \citep{ALEO2022101846}.

To address these gaps, we present \texttt{AppleCiDEr} (Applying multimodal learning to Classify transient Detections Early), a unified multimodal ML framework for time-domain classification. It is built to serve two synergistic purposes: (1) a production-ready classifier that fuses photometry, image cutouts, and metadata for early classification of both transients and variable sources in ZTF and LSST pipelines; and (2) a foundation model architecture incorporating spectra to enable archival reanalysis and cross-domain generalization.

Section~\ref{sec:data} describes the data and preprocessing. Each modalities model is presented separately: photometry (Sec.~\ref{sec:photometry}), images/metadata (which we treat as a single modality due to their presence in the alert packets; Sec.~\ref{sec:images}), and spectra (Sec.~\ref{sec:spectra}). Sec.~\ref{sec:disc} summarizes the results and discusses the future plans for \texttt{AppleCiDEr}. 

The \texttt{AppleCiDEr} pipeline is publicly available at \url{https://github.com/skyportal/applecider}. We release the software and training set we used from the Bright Transient Survey and full ZTF ID listing for reproducibility.

\section{Dataset} \label{sec:data}

Alert and photometry data from ZTF are stored in \texttt{Kowalski}\footnote{\url{https://github.com/skyportal/Kowalski}}, an open-source, multi-survey data archive and alert broker \citep{duev2019}, from which we query these modalities. For sources with spectra, we gather them from \texttt{Fritz}\footnote{\url{https://github.com/fritz-marshal/fritz}}, the version of \texttt{SkyPortal} \citep{vanderWalt2019, 2023ApJS..267...31C} used in production by the ZTF collaboration. Further spectra were gathered from public data repositories including WISeREP \citep{Yaron_2012}, SDSS \citep{2022ApJS..259...35A}, DESI \citep{2024AJ....168...58D}, as well as the GROWTH Marshal \citep{Kasliwal_2019} and Transient Name Server \footnote{\url{https://www.wis-tns.org/}}.

All ZTF sources in the dataset\footnote{Complete source list: \url{https://github.com/skyportal/applecider}} were classified into one of the transient categories (Table \ref{tab:unified_distribution}) 
with probability $\geq 0.6$ using non-machine learning methods. Sources were selected regardless of magnitude, though half originate from the Bright Transient Survey \citep{2020ApJ...895...32F, Perley_2020, Rehemtulla+2024}. In the following, we describe the pre-processing steps for each modality.

\begin{table}[htbp]
\centering
\caption{Unified Class Distribution by Object Type and Dataset. Note that 9376 out of 18245 objects are from BTS.}
\label{tab:unified_distribution}
\scriptsize
\begin{tabular}{l|rr|rr}
\toprule
\toprule
Type & Number of objects \\
\midrule
SN Ia  & 6322 \\
SN Ib  & 145  \\
SN Ic  & 153\\
SN II\textsuperscript{a}  & 695  \\
SN IIb & 126   \\
SN IIP & 627  \\
SN IIn & 234  \\
CV\textsuperscript{c} & 963 \\
AGN\textsuperscript{d}  & 8905  \\
TDE & 75 \\
\midrule
\textbf{Total} & \textbf{18245} \\
\bottomrule
\end{tabular}
\\
\textsuperscript{*}SN I: Type I Supernovae,
\textsuperscript{a}Type II Supernovae,
\textsuperscript{b} Cataclysmic Variables,
\textsuperscript{c} Active Galactic Nuclei
\end{table}

\subsection{Alerts}

Alerts are triggered by a $ \geq 5\sigma$ change in brightness relative to historical reference images. These events arise from genuine astrophysical sources (e.g., transients, variable stars, or moving objects like asteroids) or artifacts \citep{2019PASP..131g8001G}. Each alert packet contains contextual information, including metadata features and a triplet of $63 \times 63$ pixel image cutouts centered on the transient candidate:
\textit{(i) a science image} (new observation),
\textit{(ii) a reference image} (co-add of historical data), and
\textit{(iii) a difference image} (subtraction result). Multiple alerts may be generated per object over the survey lifetime, reflecting distinct activity epochs \citep{2019PASP..131a8003M}. Alert data include photometry in the ZTF-$g$, ZTF-$r$, and ZTF-$i$ filters \citep{2020PASP..132c8001D}, which are converted from magnitudes to fluxes. 

While alerts do not include spectroscopic information, we incorporate such data when available for specific objects (for details, see Sec. \ref{sec:preprocess_spectra}). A unified class Distribution by object type and dataset is given in Table \ref{tab:unified_distribution}.

Our analysis includes 17 features, which we include based on experience from \texttt{BTSBot} \citep{Rehemtulla+2024}, as listed in Table \ref{table:meta}. For a complete description of the alert metadata, we refer the reader to the \textit{ZTF Alert Schema}\footnote{\url{https://zwickytransientfacility.github.io/ztf-avro-alert/schema.html}}. Additional details on the photometric data are provided in Sec.~\ref{sec:photometry}. 

\begin{table}[htbp]
\scriptsize 
\centering
\caption{\label{table:meta} Overview of metadata features adopted from \texttt{BTSBot}.}
\begin{tabular}{p{3cm} p{5cm}}
\hhline{==}
\textbf{Feature name} & \textbf{Description [unit]} \\
\hline
\multicolumn{2}{c}{\textbf{Alert Metadata}} \\
\hline
\texttt{sgscore\{1,2\}} & Star/galaxy classification score for two closest PS1 objects \\
\texttt{distpsnr\{1,2\}} & Angular separation to two closest PS1 sources [arcsec] \\
\texttt{ra} & Right Ascension [deg]\\
\texttt{dec} & Declination [deg] \\
\texttt{magpsf} & Magnitude from PSF-fit photometry [mag] \\ 
\texttt{sigmapsf} & $1\sigma$ uncertainty in \texttt{magpsf} [mag] \\ 
\texttt{chinr} & $\chi$ parameter of nearest source in reference image PSF-catalog \\  
\texttt{sharpnr} & Shape measurement of nearest reference object \\ 
\texttt{sky} & Background sky brightness [data units] \\
\texttt{classtar} & Star/Galaxy classification score from SExtractor \\
\texttt{ndethist} & Spatially-coincident detections within 1.5 arcsec  \\ 
\texttt{ncovhist} & Times candidate position on any field/readout-channel \\
\texttt{scorr} & Peak pixel signal-to-noise ratio in detection \\ 
\texttt{nmtchps} & Source matches from PS1 catalog within 30 arcsec \\
\texttt{diffmaglim} & Magnitude for $5\sigma$ detection [mag] \\ 
\hline
\multicolumn{2}{c}{\textbf{Custom Metadata}} \\
\hline
\texttt{age} & Total duration: \texttt{days\_since\_peak + days\_to\_peak} \\
\texttt{nnondet}\textsuperscript{a}  & Difference: \texttt{ncovhist - ndethist} (Non-detections) \\
\hline
\end{tabular}
\textbf{Notes.} PSF = Point Spread Function. S/N = Signal-to-noise ratio.\\
\textsuperscript{a}Derived following \citet{CarrascoDavis2021}
\end{table}

\subsection{Spectra}
\label{sec:preprocess_spectra}

Although alert packets themselves do not contain spectroscopic data, we associate each alert with a single archival spectrum when available. These spectra are primarily obtained with the Spectral Energy Distribution Machine (SEDM; \citealt{Blagorodnova_2018, 2019A&A...627A.115R, 2022PASP..134b4505K}) with additional observations contributed by various other instruments (for example, see Table~\ref{table:inst}). For consistency, we attach the spectrum that is temporally closest to the object's first photometric observation for the multimodal model, while the spectral-only model is trained using all available spectra per object to capture spectral evolution.
Spectral labels are drawn from transient classifications provided by the Transient Name Server (TNS) or derived from internal classification pipelines.

Each spectrum undergoes a standardized preprocessing pipeline designed to restore intrinsic spectral features and produce input suitable for deep learning models. The first step involves correcting for redshift. Given a spectroscopic redshift $z$, we transform the observed wavelength axis to the rest-frame using
\begin{equation}
    \lambda_{\text{rest}} = \frac{\lambda_{\text{obs}}}{1 + z}.
\end{equation}

We then restrict all spectra to the rest-frame wavelength range of $3850\,\text{\AA}$ to $8500\,\text{\AA}$. Spectra are linearly interpolated to a fixed grid of 4096 evenly spaced points within this range. Spectra that do not fully cover the target wavelength range yield undefined values at certain grid points during interpolation; we set these regions to zero to indicate the absence of valid data.
The resulting flux arrays are then standardized using Z-score normalization:
\begin{equation}
    f_{\text{norm}} = \frac{f - \mu}{\sigma},
\end{equation}
where $\mu$ and $\sigma$ are the mean and standard deviation computed from the valid (nonzero) flux values of each spectrum.

This procedure ensures that all spectra are aligned in rest-frame wavelength, resampled to a fixed resolution, and normalized in scale, providing consistent and model-friendly inputs.

\begin{table}[htbp]
\centering
\caption{Spectroscopic observations by instrument.}
\label{table:inst}
\setlength\tabcolsep{8pt}
\renewcommand{\arraystretch}{1.2}
\begin{tabularx}{\linewidth}{l r}
\hline
\hline
\textbf{Spectrograph (or Survey)} & \textbf{Number of Spectra} \\
\midrule
SEDM\textsuperscript{1} & 5952 \\
SDSS\textsuperscript{2} & 2980 \\
SPRAT\textsuperscript{3} & 308 \\
DBSP \textsuperscript{4} & 303 \\
EFOSC2\textsuperscript{5} & 196 \\
DESI\textsuperscript{6} & 158\\
LRIS\textsuperscript{7} & 145\\
SNIFS \textsuperscript{8} & 125\\
DeVeny &  79\\
FLOYDS-N, FLOYDS-S & 63\\
ALFOSC \textsuperscript{9} & 62 \\
DIS \textsuperscript{10} & 56\\
Goodman & 24\\
AFOSC\textsuperscript{11} & 17\\
Gemini-N, Gemini-S & 12\\
Other Spectrographs \textsuperscript{12} & 76 \\

\bottomrule
\end{tabularx}

\vspace{1em}
\textsuperscript{1}{Spectral Energy Distribution Machine (SEDM)}
\textsuperscript{2}{Sloan Digital Sky Survey (SDSS)}
\textsuperscript{3}{SPectrograph for the Rapid Acquisition of Transients (SPRAT)}
\textsuperscript{4}{Double Spectrograph (DBSP)}
\textsuperscript{5}{ESO Faint Object Spectrograph and Camera (EFOSC2)}
\textsuperscript{6}{Dark Energy Spectroscopic Instrument (DESI)}
\textsuperscript{7}{Low-Resolution Imaging Spectrograph (LRIS)}
\textsuperscript{8}{SuperNova Integral Field Spectrograph (SNIFS)}

\textsuperscript{9}{Alhambra Faint Object Spectrograph and Camera (ALFOSC)}

\textsuperscript{10}{Dual Imaging Spectrograph (DIS)}

\textsuperscript{11}{Asiago Faint Objects Spectrograph and Camera (AFOSC)}
\textsuperscript{12}{Spectra from spectrographs with less than 10 counts}

\end{table}

\section{Photometry} \label{sec:photometry}

Early work in photometric time-series classification focused on extracting handcrafted features \citep{fats, Richards_2011} from multi-band light curves—such as parametric fits \citep{Karpenka2012, Lochner2016, Noebauer2017, Villar2019}, principal component analyses \citep{Ishida2013, Lochner2016}, and wavelet or Gaussian process models \citep{Varughese2015,Boone2019}. Class imbalance and limited coverage of rare transients made simulated datasets like the \texttt{SNPhotCC} \citep{Kessler2010} and \texttt{PLAsTiCC} key benchmarks. 

As data volumes have grown, deep learning has become increasingly dominant for machine-learning based classifications. Specifically, Transformer models have shown state-of-the-art performance on both synthetic and real photometric datasets, with architectures such as \texttt{TimeModAttn} \citep{Pimentel2023} and multimodal models like \texttt{ATAT} \citep{CabreraVives2024} and \texttt{AstroM}$^3$ leveraging attention to integrate photometry with metadata and spectra.

Motivated by these advances, we adopt two attention-based encoders tailored to ZTF light curves of explosive transients. The first is the Informer \citep{Informer}, which employs sparse attention to efficiently model long sequences and has proven effective in \texttt{AstroM}$^3$. The second, a dense Transformer with a learnable classification token 
\citep[\texttt{[CLS]}-Transformer;][]
{Devlin2019, Zerveas2021}, is designed for early-epoch classification under limited observations. This architecture is enhanced with learnable temporal embeddings and trained in two stages: self-supervised pretraining and supervised fine-tuning. Below, we present key information about the input data, model architecture, and training. For detailed discussion on these aspects, data preprocessing and ablation study comparison with other models, readers are referred to \cite{Felipe2025}.

\subsection{Input Representation}
Raw photometric observations $(t_i, f_i, \sigma_{f,i}, b_i)$ are first grouped into events. Measurements within a 12\,hr window are temporally merged using inverse-variance weighting, resulting in a set of effective observations. 

Each merged observation is encoded as a 7-dimensional feature vector designed to capture both photometric properties and temporal context (Table \ref{tab:input_features}). The temporal features ($\Delta t_n$, $\delta t_n$) explicitly encode absolute and relative timing information, addressing the irregular sampling inherent in survey data. Log-scaled flux with uncertainty propagation ensures numerical stability while preserving physical interpretability. Filter encoding and color availability flags enable the model to handle multi-band observations and adapt to varying photometric coverage.

\begin{deluxetable}{ll}
\tabletypesize{\small}
\tablecaption{Input Features for Attention-Based Classification\label{tab:input_features}}
\tablehead{
\colhead{\textbf{Feature}} & \colhead{\textbf{Description}}
}
\startdata
$\Delta t_n$ & Time since first detection ($t_n - t_1$) \\
$\delta t_n$ & Time since previous observation ($t_n - t_{n-1}$) \\
$\log_{10}(f_n)$ & Log-scaled flux with floor thresholding \\
$\sigma_{\log f,n}$ & Uncertainty on log-flux (propagated from $\sigma_{f,n}$) \\
$\boldsymbol{b}$ & One-hot encoding vector for $g$, $r$, or $i$ band \\
\enddata
\end{deluxetable}

\subsection{Model Architecture}

Our \texttt{[CLS]}-Transformer architecture addresses three key requirements for transient classification: handling irregular temporal sampling, integrating multi-band photometry, and enabling early classification from partial light curves.

\paragraph{Temporal Embedding.}
Rather than fixed positional encodings, we employ learnable temporal embeddings that explicitly model the continuous nature of astronomical time series. The temporal encoding for observation $n$ combines linear and sinusoidal components. This formulation captures both monotonic evolution (linear trend) and potential periodic behavior (sinusoidal terms) while adapting to the specific temporal scales present in the training data.

\paragraph{Feature Projection and Sequence Construction.}
Each preprocessed observation vector $\tilde{x}_n$ is projected into the model's embedding dimension through a learned transformation. The final input representation combines photometric features with temporal context. A special classification token \texttt{[CLS]} is prepended to each sequence, providing a dedicated representation for downstream classification that can aggregate information across all observations.

\paragraph{Transformer Encoder.}
The encoder consists of $L$ standard Transformer layers with multi-head self-attention and feedforward sublayers. The self-attention mechanism enables each observation to attend to all others in the sequence, naturally handling irregular sampling and discovering relevant temporal relationships for classification.
The final \texttt{[CLS]} representation is normalized and passed through a classification head.

\subsection{Training Procedure}
The performance of the model is evaluated using a horizon sweep approach with multiple metrics: macro-average precision recall (Macro-AUPRC), balanced precision, and Top-k inclusion rates. These metrics quantify classification performance across different observational depths and class distributions. 

A two-stage training procedure is employed, designed to leverage both unlabeled photometric data for light curve reconstruction and labeled data for transient classification.

\paragraph{Self-Supervised Pretraining.}
To initialize the encoder with light-curve-specific structure, the model is first trained using masked event modeling on light curves spanning 100 days since the first trigger. Roughly 30\% of observations are masked, and the model is trained to reconstruct key photometric features using a combination of three losses -- a mean squared error loss on the flux values, a cross-entropy loss on the photometric band labels, and a mean squared error loss on the relative timing between observations. Each of these components is weighted by a corresponding hyperparameter to balance their contributions.

\paragraph{Supervised Fine-tuning.}
For classification, the pretrained model is fine-tuned using truncated light curves (e.g., up to 100 days). We use weighted focal loss \citep{lin2018focallossdenseobject} and augmentation of the TDE class to handle the imbalance inherent to the dataset.
Two fine-tuning strategies are considered: \textit{(i)} full-model training from the start; and \textit{(ii)} training only the classification head initially, followed by gradual unfreezing of the encoder layers with lower learning rates.

\subsection{Results}

We evaluate the classification performance of our adopted model in comparison to the \texttt{Informer} encoder used in \texttt{AstroM}$^3$, both fine-tuned on light curves cut-off at 30 days since the first observation. As shown in Table~\ref{tab:classification_results_photometry}, our model achieves a significantly higher overall accuracy of 88\% compared to 59\% from the \texttt{Informer} baseline. 

\begin{figure}
    \centering
    \includegraphics[scale=0.6]{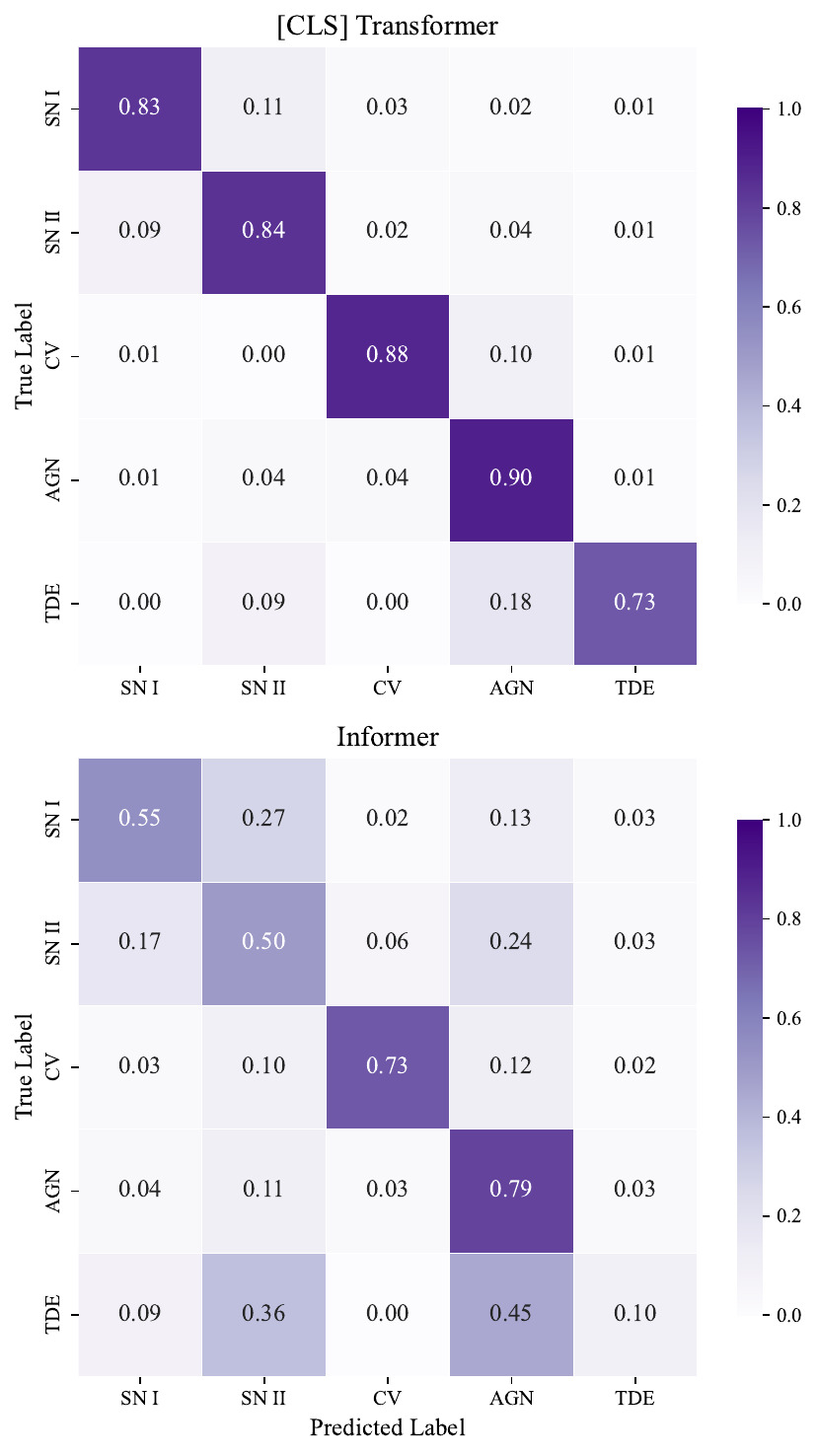}
    \caption{Confusion matrix comparison between our proposed Transformer encoder (\textit{top}) and the Informer encoder from \texttt{AstroM}$^3$ (\textit{bottom}).}
    \label{fig:confusion matrix photometry}
\end{figure}

Our model shows consistent improvements across all transient classes, and better per-class AUC scores. It excels with challenging classes like TDEs, achieving 0.95 AUC versus 0.82. The confusion matrix in Fig.~\ref{fig:confusion matrix photometry} highlights our model's enhanced class separability and reduced ambiguity.

\begin{table}[htbp]
\centering
\caption{Classification Performance Comparison for Photometry only.}
\label{tab:classification_results_photometry}
\begin{tabular}{l|c|c}
\hline
\hline
\textbf{Model} & \textbf{\texttt{Informer}} & \textbf{\texttt{[CLS] Transformer}} \\
\hline
\textbf{Overall Accuracy (\%)} & 59 $\pm$ 0.3 & \textbf{88 $\pm$ 0.4} \\
\hline
\multicolumn{3}{c}{\textbf{AUC Values by Class}} \\
\hline
Macro-average & 0.89 & \textbf{0.97} \\

SN I & 0.94 & \textbf{0.99} \\

SN II & 0.87 & \textbf{0.96} \\

CV & 0.90 & \textbf{0.97} \\

AGN & 0.95 & \textbf{0.99} \\

TDE & 0.82 & \textbf{0.95} \\

\bottomrule
\end{tabular}
\end{table}

\section{Images: \texttt{A}\MakeLowercase{\texttt{stro}}\texttt{M}\MakeLowercase{\texttt{i}}\texttt{NN}} \label{sec:images}

Traditional astronomical transient classification relied on feature-based machine learning models 
\citep{bloom2012, brink2013}, including photometric redshift estimation \citep{carrasco2013} 
and supernova classification \citep{villar2020}. The advent of CNNs revolutionized image processing, enabling morphology classification \citep{dieleman2015}, 
strong lens detection \citep{lanusse2018}, and real-bogus filtering \citep{duev2019}. Recent multi-input architectures \citep{CarrascoDavis2021, duev2021phenomenologicalclassificationzwickytransient} combine imaging and tabular data, 
directly inspiring \texttt{BTSbot}'s \citep{Rehemtulla+2024} design. 

\texttt{BTSbot} utilizes a multi-input CNN combining images and metadata features (i.e., see Tab.~\ref{table:meta}) to identify bright extragalactic transients and infant supernovae \citep{Rehemtulla+2025} in ZTF. This approach processes images and features separately and then combines them to improve classification accuracy over single-input methods. The extracted features are processed by a multi-layer perceptron (MLP; \citealt{rumelhart1986}), 
while image data is handled by a convolutional branch.  

The existing methods of analyzing images and metadata only use two MLPs, one for the metadata, and another to fuse the image and metadata features. While this
achieves relatively good results for large datasets and roughly even class distributions, the performance decreases in cases of uneven class distribution and limited datasets. To tackle this challenge, we introduce \texttt{AstroMiNN}, a specialized multimodal neural network that integrates convolutional vision backbones with structured expert towers for metadata data. 

\begin{figure*}[ht]
\centering
\includegraphics[width=0.9\linewidth]{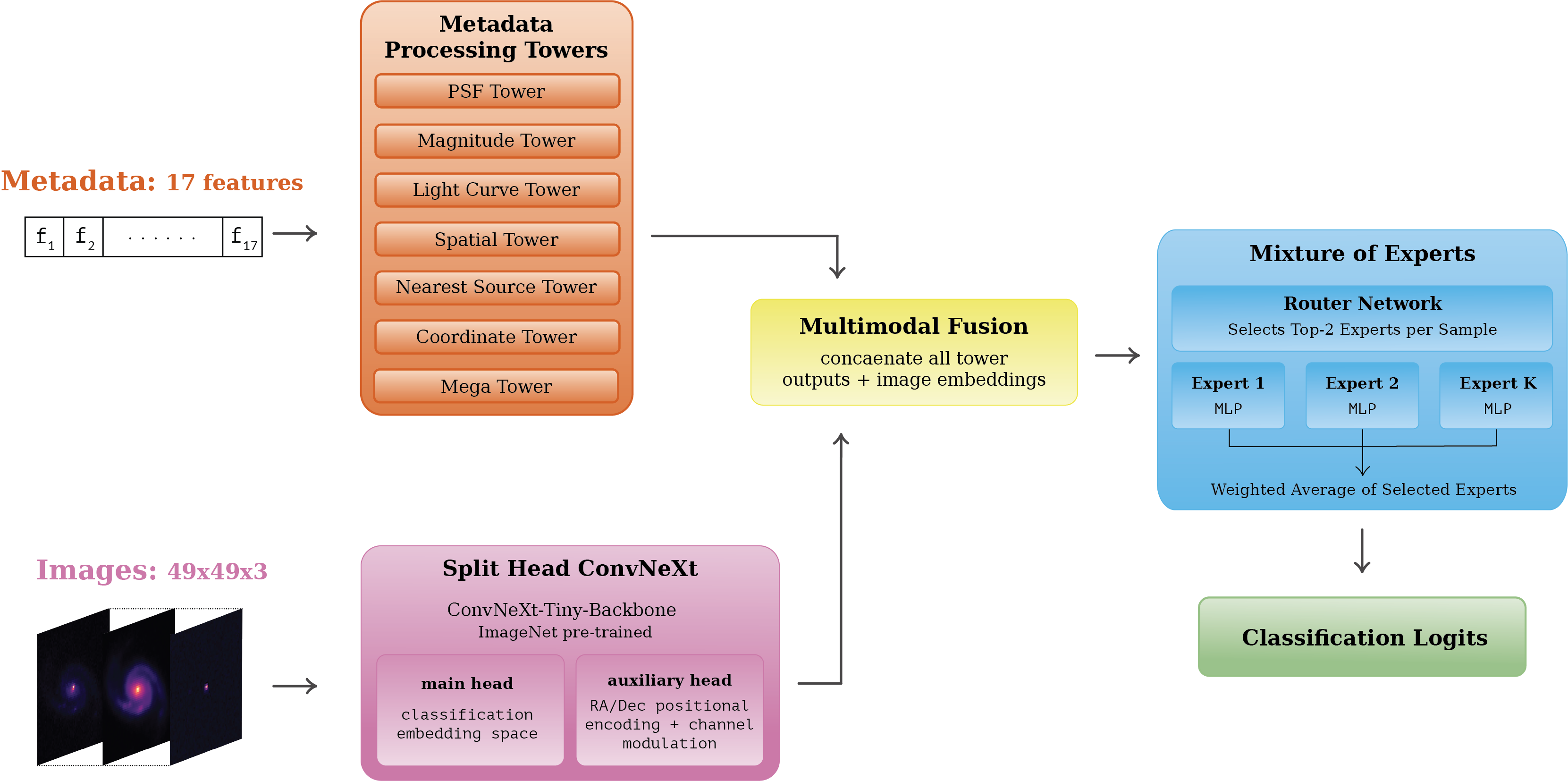}
\caption{\textbf{AstroMiNN Architecture.} Metadata features are routed through specialized residual towers before being fused with image embeddings from a dual-head \texttt{ConvNeXt}-based encoder. The combined representation is processed by a Mixture-of-Experts (MoE) module, with expert selection governed by a router network.}
\label{fig:XastroMiNN}
\end{figure*}

The \texttt{AstroMiNN} architecture is explicitly designed to incorporate domain-specific structure, leveraging both the spatial and physical properties of transient detections. The overall architecture is illustrated in Figure~\ref{fig:XastroMiNN} and for each part of the architecture, we give more details in the following.

\subsection{Metadata Processing Towers}

The metadata stream is decomposed into thematic groups and processed by independent \emph{Residual Tower Blocks} \citep{He2016}. Each tower captures distinct scientific parameters:

\begin{itemize}
\item \textbf{PSF Tower:} Shape parameters and PSF sharpness (e.g., \texttt{chinr}, \texttt{sharpnr}, and \texttt{classtar}).
\item \textbf{Magnitude Tower:} Photometric statistics such as PSF magnitude and peak magnitude so far (e.g., \texttt{magpsf}, \texttt{sigmapsf}, and \texttt{diffmaglim}).
\item \textbf{Light Curve Tower:} Historical variability and timing-based metrics (e.g., \texttt{age}, \texttt{ndethist}, \texttt{ncovhist}, and \texttt{nnondet}).
\item \textbf{Spatial Tower:} Spatial observables including sharpness and source confusion (e.g., \texttt{sky}, \texttt{nmtchps}, \texttt{scorr}).
\item \textbf{Nearest Source Towers:} Proximity and likelihood features from the nearest detected source (\texttt{sgscore} and \texttt{distpsnr} pairs).
\item \textbf{Coordinate Tower:} Right Ascension (\texttt{ra}) and Declination (\texttt{dec}), optionally modulated with sinusoidal encodings.
\item \textbf{Mega Tower:} A comprehensive tower trained on all metadata features jointly.
\end{itemize}

Each tower is composed of a small feedforward block with \textbf{GELU activations} \citep{Hendrycks2016} and residual connections \citep{He2016}. Optional \textbf{gating heads} allow for dynamic feature masking based on learned confidence scores \citep{Shazeer2017}. 

We construct towers in this way such that each metadata feature is considered primarily in relation to a select few other related metadata items. This approach requires that each metadata feature is first expanded with its most directly associated features before forming more distant connections. For example, magnitude measurements should be expanded with their associated error values before establishing further feature linkages, thus enforcing any necessary conditionality among related features. Given that these are merely initial expansions, each tower is designed to be lightweight, incorporating only two linear expansion layers and a skip path, all maintaining the same-dimensional output.

\subsection{Image Feature Encoder}

For the imaging stream, we adopt a \texttt{ConvNeXt-Tiny} backbone \citep{liu2022convnet2020s} pre-trained on \texttt{ImageNet} \citep{imagenet, hayakawa2021neural}. Our encoder, termed ``Split Head ConvNeXt'', incorporates two specialized custom heads to process the images. The main head projects features into a classification embedding space, providing the primary pathway for feature extraction and representation learning. Complementing this, the \textbf{auxiliary head} performs dynamic channel modulation using a per-sample sine-based positional embedding that incorporates RA/Dec coordinates, allowing the model to spatially contextualize each observation within the celestial coordinate system.

The input images consist of triplets of 49 $\times$ 49 pixel cutouts \citep[see Sec.~\ref{sec:data};][]{Rehemtulla+2024, CarrascoDavis2021}. Beyond these standard channels, the model processes additional information that encodes the radial distance from the cutout center and the embedded sky coordinates, enriching the spatial and positional context available to the network during feature extraction and classification.

\subsection{Fusion and Expert Routing}

While tower gating mechanisms (e.g., softmax- or sigmoid-weighted sums) have been explored before in \texttt{astro-ML} work \citep{astroML, astroMLText}, they often rely on the strong assumption that individual tower features are linearly combinable into a global representation. Instead, we adopt a different approach; tower outputs are concatenated \citep{Kiela2019} and passed to a shared fusion space without weighting. This strategy enables the network to learn richer feature interactions through multiple fusion experts -- each implemented as an independent MLP -- that operate over the full multimodal embedding.

Subsequently, the unified representation is processed by a Mixture of Experts (MoE) module \citep{Shazeer2017,Fedus2022,mu2025}. This consists of \textit{(i)} a router network that computes a probability distribution over $K$ expert heads and selects the top-2 experts per sample, and \textit{(ii)} a set of $K$ expert MLPs, each acting as a lightweight classifier. Expert load balancing follows \cite{Lepikhin2020}.

This avoids forcing towers into a globally interpretable weighting scheme and instead allows each expert to adaptively specialize on complex feature interactions. The final output logits are computed as a soft aggregation of the top-k expert predictions, promoting competition among experts while maintaining interpretability through expert load monitoring.

\subsection{Loss Function and Expert Specialization}

In addition to traditional classification losses such as Focal \citep{lin2018focallossdenseobject}, we apply a regularization term that encourages \emph{expert specialization} \citep{Eigen2014, mu2025}. The router is trained to align expert weights with class-specific priors using a target expert assignment (e.g., class label modulo number of experts). This is enforced via an auxiliary MSE loss on expert weights:
\begin{equation}
\mathcal{L}_{\text{total}} = \mathcal{L}_{\text{focal}} + \lambda \cdot \text{MSE}(w_{\text{expert}}, w_{\text{ideal}})
\end{equation}
where $\lambda$ controls the contribution of the specialization term.

\subsection{Results}
\begin{figure}
    \centering
    \includegraphics[scale=0.65]{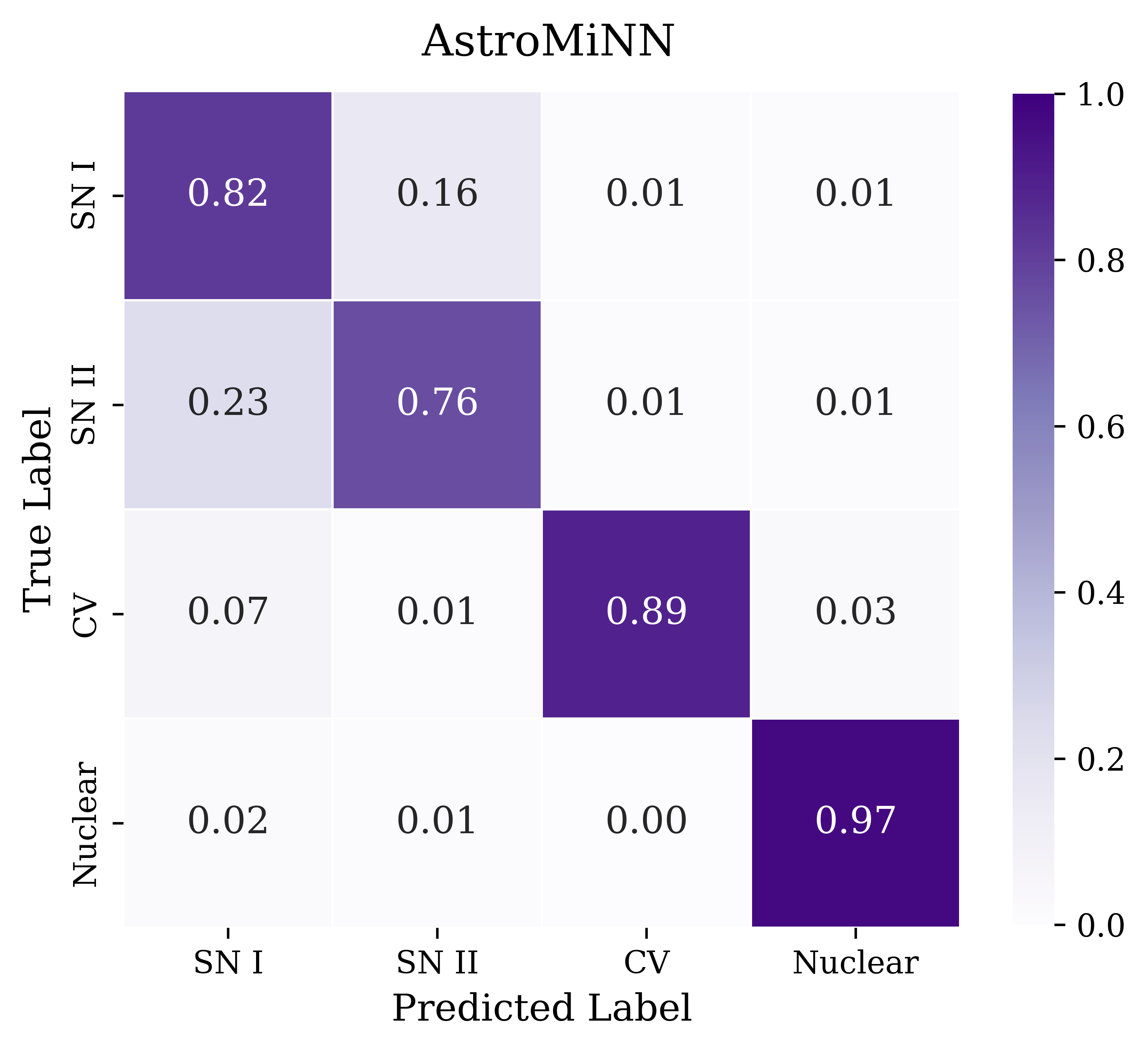}
    \caption{Confusion matrix for \texttt{AstroMiNN} model.}
    \label{fig:confusion matrix images}
\end{figure}

 After evaluating various models, we found that the \texttt{BTSbot} model provided the best classification results among the existing architectures we tried. Consequently, we compare our new model \texttt{AstroMiNN} with \texttt{BTSbot}. Table~\ref{tab:images_auc_comparison} displays the AUC scores for both models across four classes, indicating that \texttt{AstroMiNN} performs better classification for nuclear and CV events. The confusion matrix is given in Fig.~\ref{fig:confusion matrix images}.

\begin{table}[h]
\centering
\caption{Comparison of AUC scores between \texttt{AstroMiNN} and \texttt{BTSbot} models.}
\begin{tabular}{l|c|c}
\hline
\hline
\textbf{Class} & \textbf{\texttt{AstroMiNN}} & \textbf{\texttt{BTSbot}} \\ 
SN I & 0.83 & 0.83  \\
SN II & 0.76 & \textbf{0.78} \\
CV & \textbf{0.99} & 0.97 \\
Nuclear & \textbf{0.99} & 0.97 \\
\midrule
\end{tabular}
\label{tab:images_auc_comparison}
\end{table}

\section{Spectra: \texttt{S}\MakeLowercase{\texttt{pectra}}\texttt{N}\MakeLowercase{\texttt{et}}}
\label{sec:spectra}

Spectroscopic classification using machine learning often operates on flux-wavelength sequences that have been interpolated onto a fixed grid, allowing uniform input dimensions across a dataset. Recent advances in deep learning have shown promising results in this domain. For instance, \texttt{GalSpecNet} \citep{2024MNRAS.527.1163W} applies convolutional neural networks (CNN) to 1D spectra for transient classification. The same model has also been adopted in the \texttt{AstroM}$^3$ pipeline. Another example, \textit{SNIascore} \citep{Fremling2021SNIascore}, leverages recurrent units such as BiLSTM and GRU layers for real-time classification of low-resolution Type Ia supernova spectra. Similarly, \texttt{CCSNSCore} \citep{sharma2025ccsnscoremultiinputdeeplearning} adopts a multi-input architecture designed specifically to extract and classify spectroscopic features of core-collapse supernovae. Another approach has been presented in \cite{vision_transformers1}, where vision transformers for spectra data are used.

Although existing methods have achieved notable progress in spectral classification, several critical limitations remain. In particular, supernova classification strongly depends on localized spectral features -- such as the shape, depth, and width of absorption lines -- which often vary significantly across subtypes.Recurrent models, such as BiLSTM and GRU, are well-suited for modeling long-range dependencies. However, they may overlook important local structures critical to supernova subclass discrimination. Moreover, to reduce computational cost, RNN-based models often restrict input sequence length, which can lead to information loss and reduced classification accuracy. While convolutional architectures are naturally better at capturing local patterns, most existing 1D CNN-based models are relatively shallow and lack the expressive capacity to extract complex or multi-scale spectral features. This becomes particularly limiting when dealing with noisy observations or spectra that evolve over time.

To address these issues, the final adopted model is a deep, multi-scale 1D convolutional architecture—\texttt{SpectraNet}~\citep{Maojie2025}. This model is designed to improve the extraction of localized spectral signatures while maintaining computational efficiency and achieving higher classification performance. As shown in ~\citealt{Maojie2025}, this architecture outperforms traditional 1D CNN and Transformer baselines in supernova spectral classification. 

\begin{figure*}[t]
\centering
\includegraphics[width=0.95\linewidth]{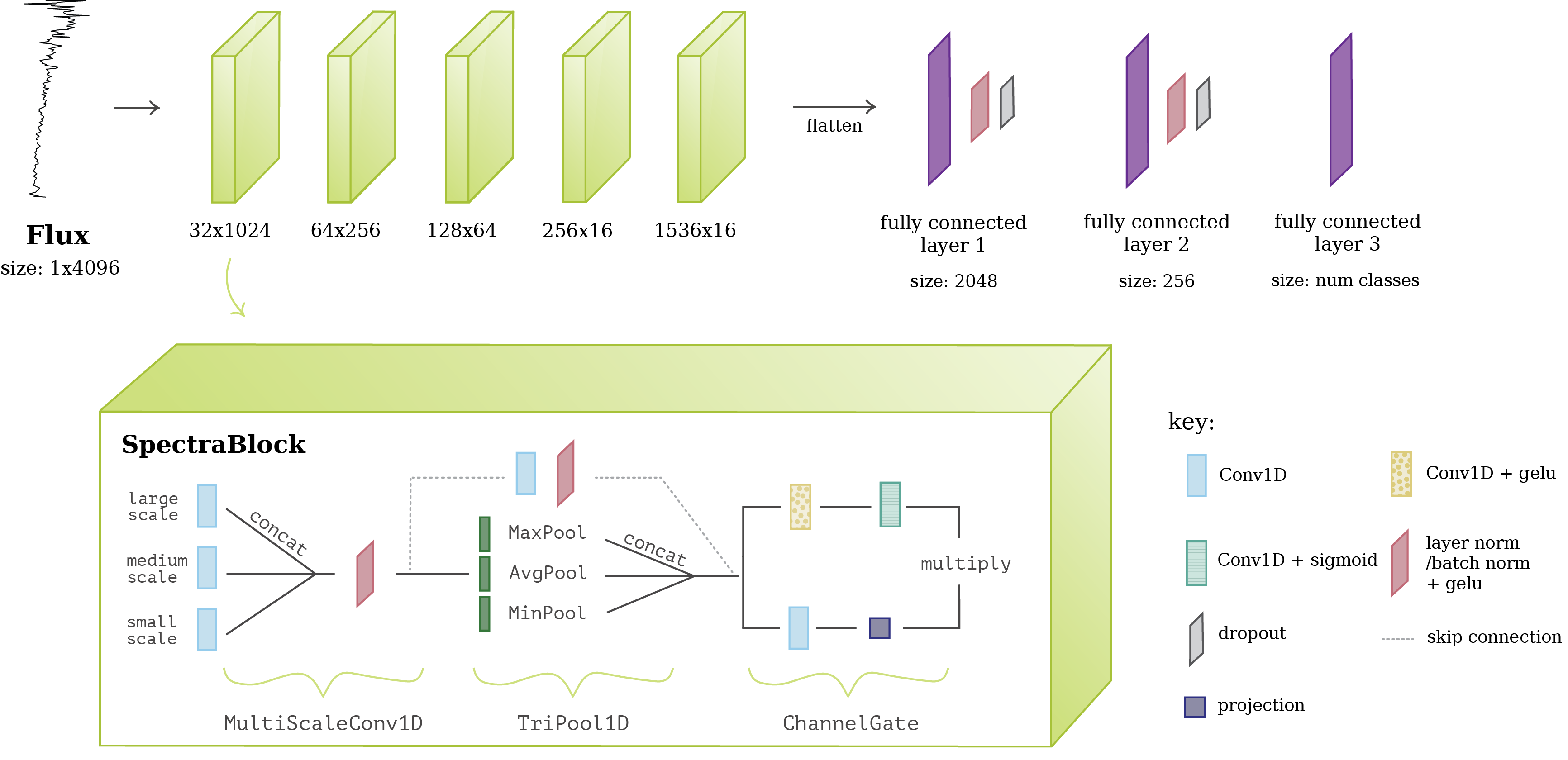}
\caption{\texttt{SpectraNet} Architecture.}
\label{fig:spectranet-diagram}
\end{figure*}

\subsection{Training Details}
The \texttt{SpectraNet}-1D \citep{Maojie2025} network consists of two main components, a convolutional feature extraction backbone followed by fully connected classification layers. 

\paragraph{Convolutional Backbone}
The network begins with five sequential \texttt{SpectraBlocks} that progressively reduce the spectral resolution while increasing the feature depth. The input spectrum (size $1 \times 4096$) is processed through \texttt{SpectraBlocks} with sizes $32 \times 1024$, $64 \times 256$, $128 \times 64$, $256 \times 16$, and $1536 \times 16$, respectively. Each \texttt{SpectraBlocks} incorporates a multi-scale convolutional architecture that processes the input at three different scales (large, medium, and small) through parallel convolutional paths. The multi-scale features are concatenated and passed through layer normalization and batch normalization before being fed into a ``TriPool1D'' module. The ``TriPool1D'' component combines three pooling strategies (MaxPool, AvgPool, and MinPool) to capture diverse statistical representations of the spectral features. This is followed by a channel gate attention mechanism that applies sigmoid activation to selectively weight the feature channels, with a projection layer and skip connection to preserve important spectral information.

\paragraph{Classification Head}
After the convolutional feature extraction, the output is flattened and processed through three fully connected layers. The first layer maps the $1536 \times 16$ feature tensor to $2048$ dimensions, followed by a second layer reducing to $256$ dimensions, and finally a classification layer outputting the number of target classes. Each fully connected layer incorporates layer normalization, batch normalization, and GELU activation functions to ensure stable training and effective feature representation.

The model is trained using the \texttt{AdamW} optimizer~\citep{loshchilov2019decoupledweightdecayregularization}, with the learning rate selected via \texttt{Optuna}~\citep{akiba2019optunanextgenerationhyperparameteroptimization}. A linear warm-up schedule is followed by a cosine annealing learning rate schedule~\citep{loshchilov2017sgdr}, and early stopping is triggered based on a composite validation score with a fixed \texttt{patience} threshold.

To address class imbalance, we use Focal Loss~\citep{lin2018focallossdenseobject} with class-balanced weighting~\citep{cui2019classbalancedlossbasedeffective}. An Exponential Moving Average \citep[EMA;][]{tarvainen2018meanteachersbetterrole} of the model weights is maintained and used for evaluation and model saving. We also adopt automatic mixed-precision (AMP) training to reduce memory usage and accelerate convergence. Model selection is based on the highest composite score on the validation set, combining accuracy, macro F1, and Top-3 accuracy.

This architecture is specifically designed for 1D spectral data, leveraging multi-scale feature extraction and attention mechanisms to capture both local spectral features and global patterns essential for accurate astronomical object classification. A diagram of the architecture is given in Fig.~\ref{fig:spectranet-diagram}.

\subsection{Results}
\begin{figure}
    \centering
    \includegraphics[scale=0.37]{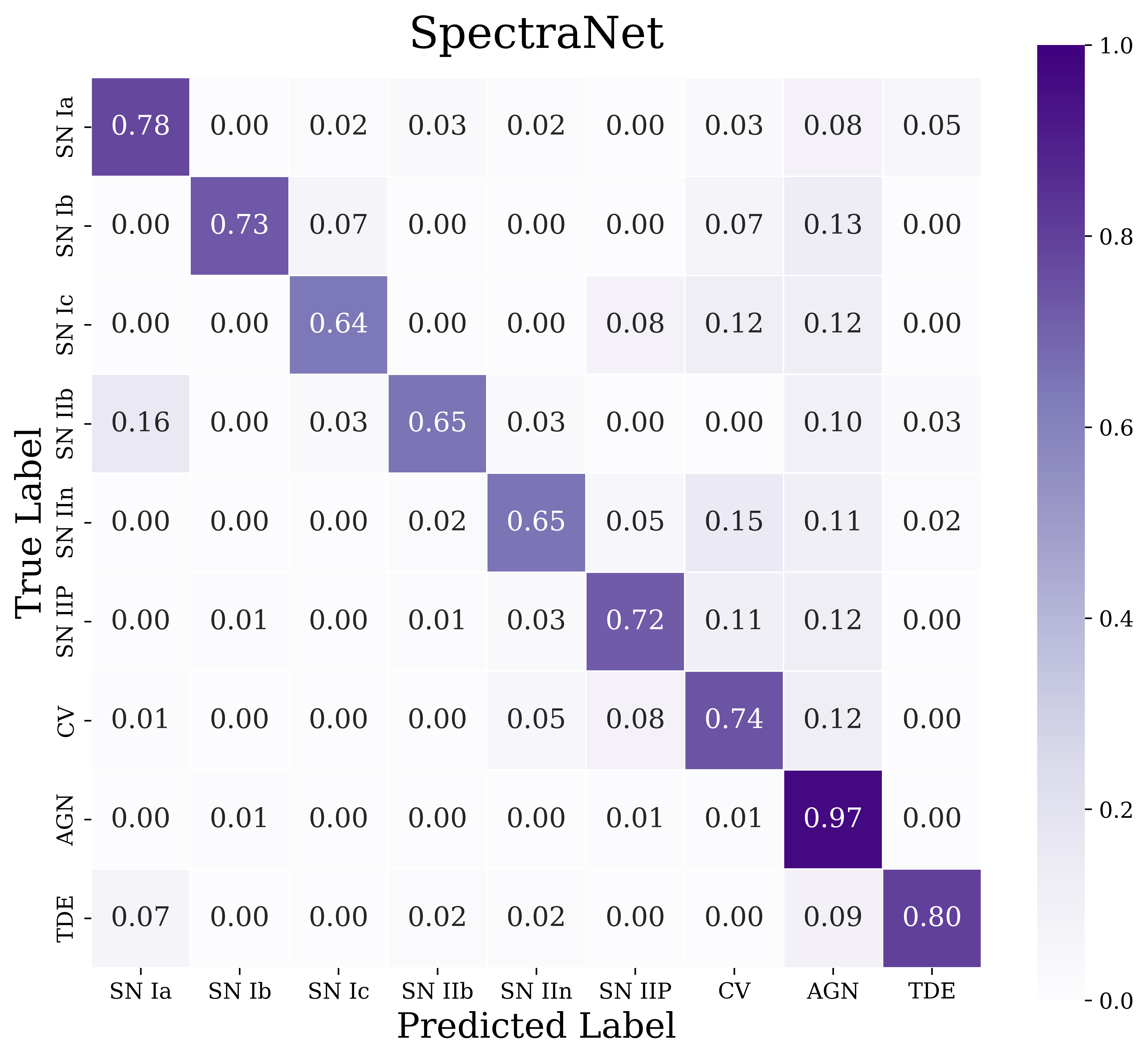}
    \caption{Confusion Matrix for \texttt{SpectraNet} model.}
    \label{fig:spectra-10-classes}
\end{figure}

Table~\ref{tab:classification_results_spectra} presents a detailed comparison of classification performance between \texttt{GalSpecNet} \citep{2024MNRAS.527.1163W} and our proposed \texttt{SpectraNet} across nine astronomical object classes. \texttt{SpectraNet} demonstrates superior overall accuracy of 87\%, outperforming \texttt{GalSpecNet}'s 83\%. In terms of class-wise AUC scores, \texttt{SpectraNet} consistently achieves higher performance across all categories. Notable improvements are observed for challenging classes such as SN Ic (AUC improved from 0.95 to 0.97), SN IIP (0.96 to 0.99), and TDE (0.96 to 0.99). Importantly, \texttt{SpectraNet} does not underperform in any category; all class-specific AUC scores are above 0.96. Figure~\ref{fig:spectra-10-classes} further illustrates the model’s confusion matrix using the \texttt{SpectraNeXt-2D} variant, highlighting substantial improvements in distinguishing core-collapse supernovae (e.g., SN IIP, IIb, IIn) and non-SN transients such as AGN, CV, and TDE.

\begin{table}[htbp]
\centering
\caption{Classification Performance Comparison}
\label{tab:classification_results_spectra}
\begin{tabular}{l|c|c}
\hline
\hline
\textbf{Model} & \textbf{\texttt{SpectraNet}} & \textbf{\texttt{GalSpecNet}} \\
\hline
\textbf{Overall Accuracy (\%)} & \textbf{87.0 $\pm$ 0.5} & 82.6 $\pm$ 0.6 \\
\hline
\multicolumn{3}{c}{\textbf{AUC Values by Class}} \\
\hline
Micro-average & \textbf{0.98} & 0.96 \\

SN Ia & \textbf{0.99} & 0.97 \\
SN Ib & \textbf{0.97} & 0.94 \\
SN Ic & \textbf{0.97} & 0.95 \\
SN IIP & \textbf{0.99} & 0.96 \\
SN IIb & \textbf{0.96} & 0.95 \\
SN IIn & 0.98 & 0.98 \\
AGN & \textbf{0.97} & 0.96 \\
CV & \textbf{0.99} & 0.97 \\
TDE & \textbf{0.99} & 0.96 \\  
\bottomrule
\end{tabular}
\end{table}

\section{ \texttt{A}\MakeLowercase{\texttt{pple}}\texttt{C}\MakeLowercase{\texttt{i}}\texttt{DE}\MakeLowercase{\texttt{r}} Performance}

\begin{table}[htbp]
\centering
\caption{Common dataset derived from objects in Tab. \ref{tab:unified_distribution}}
\label{tab:dataset}
\scriptsize
\begin{tabular}{l|rr}
\toprule
\toprule \\
Type & Objects & Alerts \\
\midrule
SN I  & 4830 & 22791 \\
SN II & 1014  & 4520  \\
CV & 279  & 1262 \\
AGN & 1251 & 2375 \\
TDE  & 36 & 117  \\
\midrule
\textbf{Total} & \textbf{7410} & \textbf{31065} \\
\bottomrule
\end{tabular}
\end{table}

\begin{figure}[t]
  \centering
  \begin{minipage}[t]{0.95\columnwidth}
    \centering
    \includegraphics[width=\textwidth]{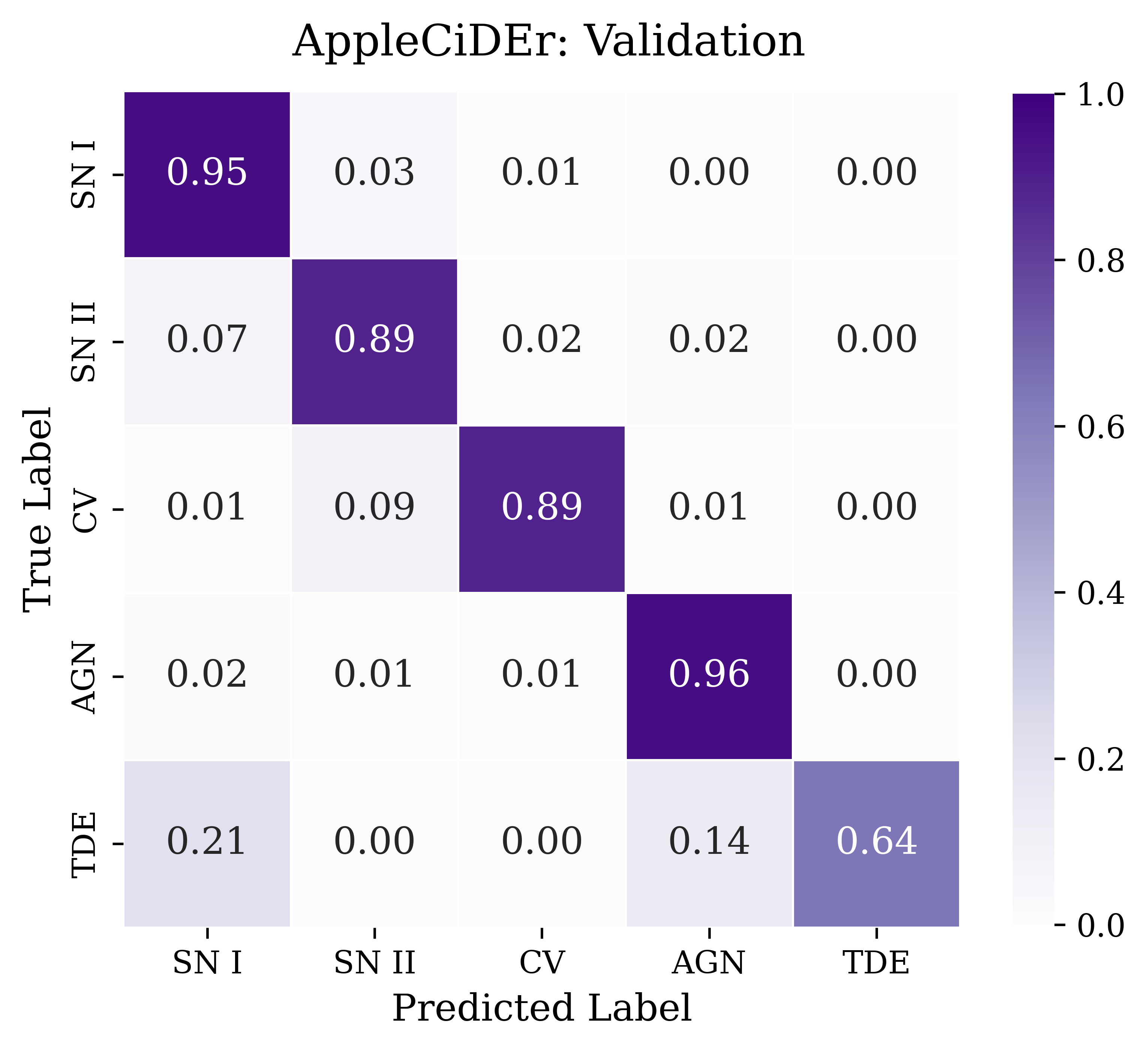}\\
    {\footnotesize (a) Validation set}
  \end{minipage}\hfill
  \\
  \begin{minipage}[t]{0.95\columnwidth}
    \centering
    \includegraphics[width=\textwidth]{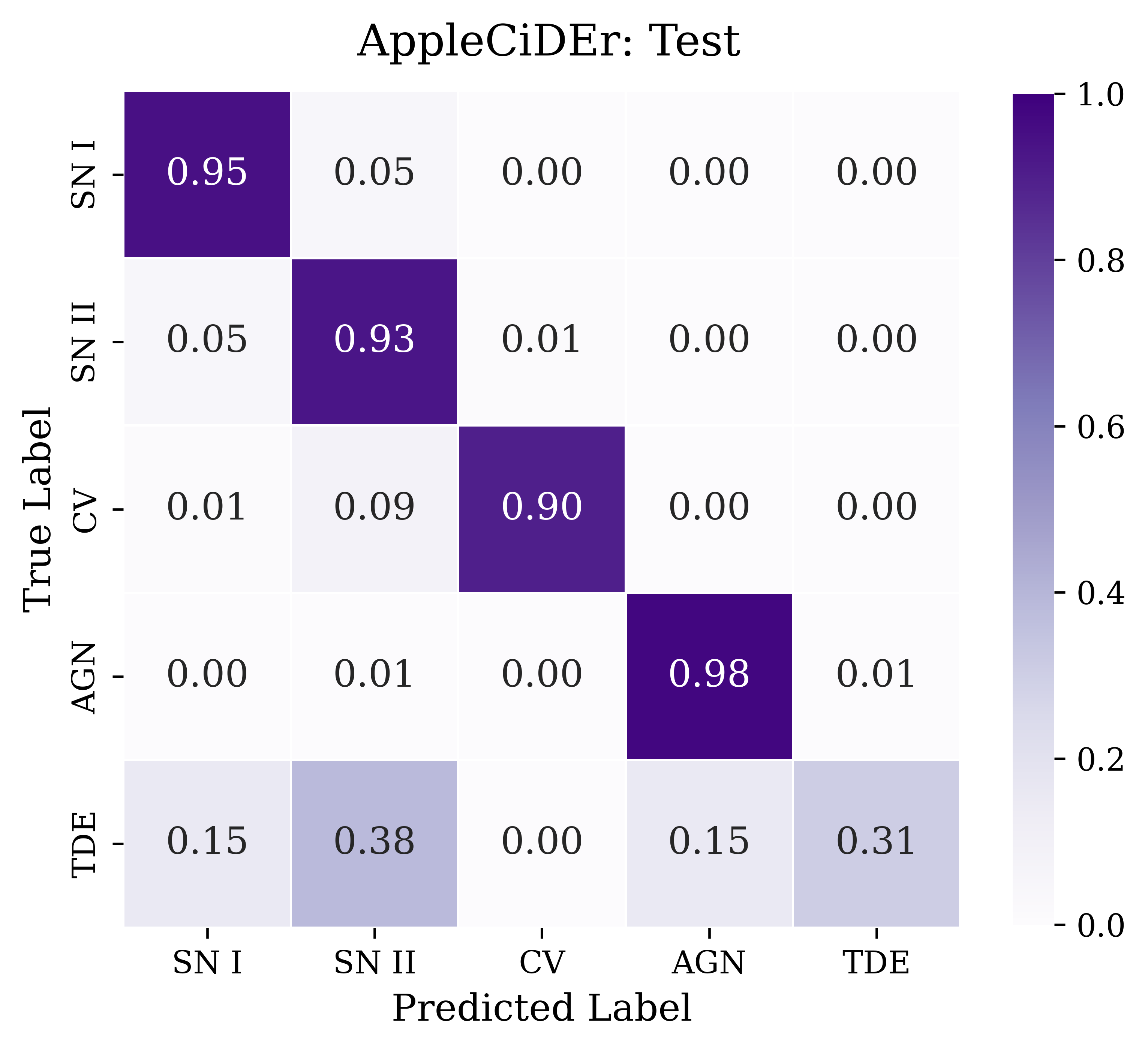}\\
    {\footnotesize (b) Test set}
  \end{minipage}
  \caption{AppleCiDEr performance. Normalized confusion matrices for the \textit{(a)} validation and \textit{(b)} test sets. The classifier shows strong performance across SN~I, SN~II, CV, and AGN classes. TDE classification is notably more challenging, with substantial confusion with SN~I, SN~II, and AGN.}
  \label{fig:cider_confmats}
\end{figure}

\texttt{AppleCiDEr} leverages the neural network architectures presented in the previous Sections~\ref{sec:photometry}-\ref{sec:spectra}, and performs class-wise averaging over their predicted probabilities. All networks are trained jointly on the same dataset -- rather than on disjoint subsets as in earlier experiments -- and are evaluated collectively on a shared test set. The dataset is chosen based on common\footnote{Available data for photometry, spectra, images and metadata.} data for ten days (see Tab.~\ref{tab:unified_distribution}). 
As shown in Fig.~\ref{fig:cider_confmats}, it demonstrates robust classification performance for most transient classes. On the validation set (Fig.~\ref{fig:cider_confmats}a), SN~I, SN~II, CV, and AGN are correctly classified with accuracies ranging from 89\% to 96\%. However, the TDE class presents challenges, with only 64\% correct identification and confusion primarily with SN~I (21\%) and AGN (14\%).

\begin{figure*}[tbp]
\centering
\includegraphics[width=0.92\textwidth]{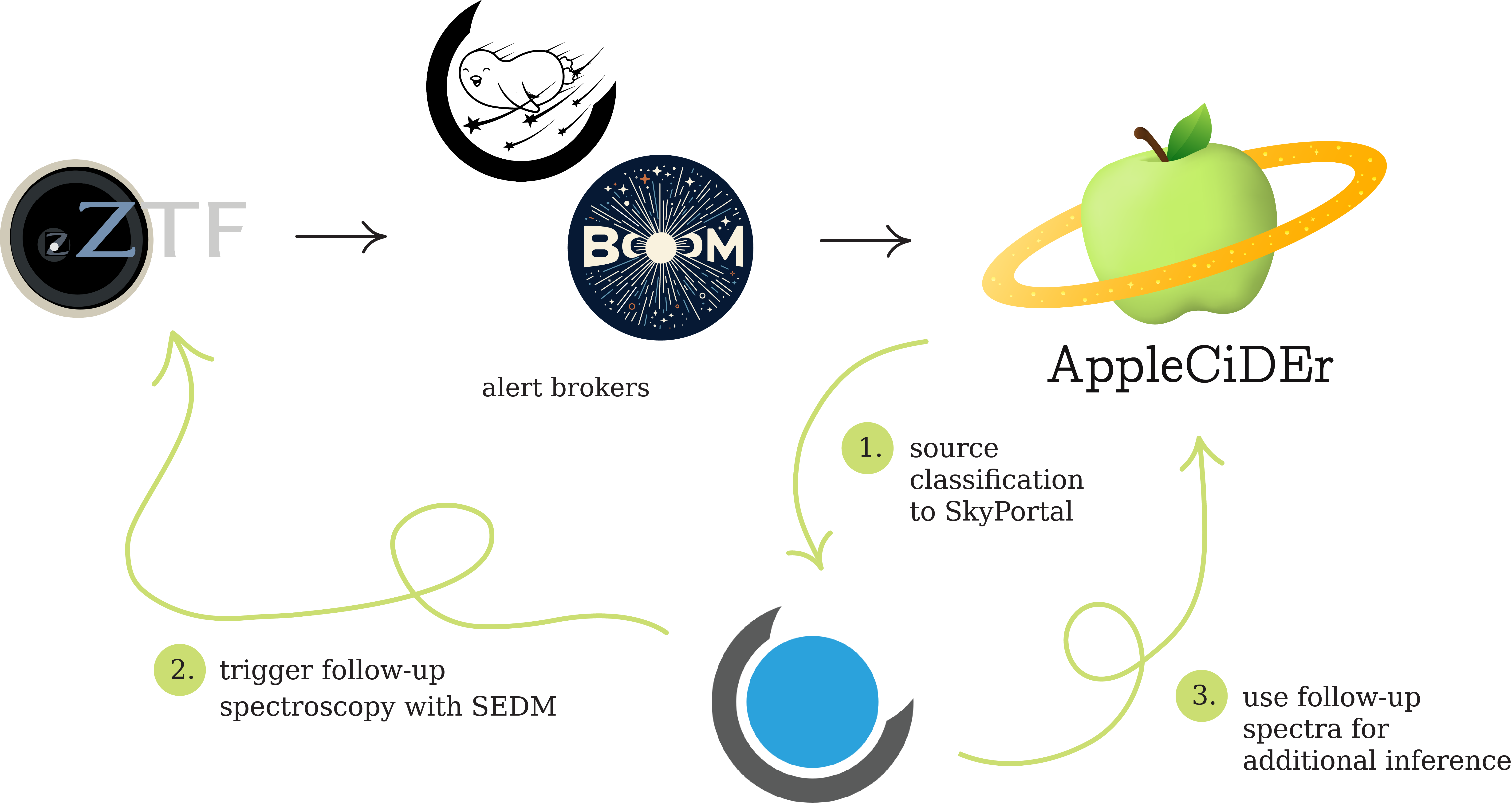}\caption{Planned ZTF production pipeline.}
\label{fig:production}
\end{figure*}

On the test set (Fig.~\ref{fig:cider_confmats}b), similar trends are observed. SN~I, SN~II, CV, and AGN remain well separated, with classification accuracies of 95\%, 93\%, 90\%, and 98\%, respectively. The TDE class exhibits a more pronounced misclassification, particularly toward SN~II (38\%) and SN~I/AGN (15\% each), achieving only 31\% accuracy.

This suggests that while \texttt{AppleCiDEr} generalizes well for common transients, TDE detection remains an area for targeted improvement. The reason for this is due to limited samples from this particular case.

\section{Discussion} \label{sec:disc}
\texttt{AppleCiDEr} introduces a unified, multimodal framework that marks a substantial advancement in time-domain astronomy. By combining photometry, image cutouts, metadata, and spectral data, the model bridges a long-standing gap in survey pipelines, which traditionally process these modalities individually or through specialized models. Prior efforts have typically focused on specific source classes -- such as supernovae or variable stars -- or have operated only on simulated datasets. In contrast, \texttt{AppleCiDEr} is designed and validated using real ZTF alert data, delivering robust and generalizable performance across a diverse range of transients and variables. This capability addresses a critical limitation in existing pipelines, which rely on modular, single-modality classifiers to process complex and heterogeneous data streams.

There are several core architectural innovations underpinning this progress. Our \texttt{[CLS]}-Transformer, adopted from \citealp{Felipe2025}), enables classification within the first few nights of observations, which is crucial for rapid response to targets such as TDEs. We also introduce a new network for image classification, the \texttt{AstroMiNN}, which leverages a metadata tower and the MoE fusion mechanism, achieving better performance compared to what exists currently. In the spectral domain, \texttt{SpectraNet-1D} \citep{Maojie2025} is adopted, which outperforms other pipelines, offering a powerful tool for low-resolution spectral classification.

These technical improvements translate directly into tangible astronomical benefits, including high classification AUCs for well-represented types such as SNIa and CVs, and up to 0.95 for more challenging classes like SNIc -- enabling more efficient early-phase identification. This, in turn, helps prioritize limited follow-up resources, especially in the context of rare transients where spectroscopic confirmation is expensive or infeasible. 

Importantly, the modularity of our approach also ensures that the framework is adaptable to upcoming large-scale surveys such as LSST, which will deliver tens of millions of nightly alerts. The auxiliary spatial head integrated into the \texttt{ConvNeXt} backbone --leveraging sky coordinates -- adds contextual awareness to image classification, a capability absent in earlier systems.

However, some limitations remain. The model continues to face challenges associated with class imbalance, where rare types like TDEs and stripped-envelope supernovae (e.g., SN~Ib/c) display increased classification uncertainty. Future versions of \texttt{AppleCiDEr} will explore fusion techniques to combine networks for each modality rather than just averaging the output, while exploring different techniques to increase the performance for the TDEs. 

In parallel, we are working on delivering this package for production. Our aim is to perform online classification with seamless integration into brokering pipelines. To this end, we are developing a real-time framework that leverages \texttt{AppleCiDEr} to classify astronomical transients from ZTF in a low-latency, automated manner; we are currently deploying \texttt{AppleCiDEr} into \texttt{SkyPortal} \citep{vanderWalt2019, 2023ApJS..267...31C} as a classification pipeline and soon will be doing so within our broker \texttt{BOOM} (Burst \& Outburst Observations Monitor) as a classification pipeline. 
As shown in Fig.~\ref{fig:production}, \texttt{AppleCiDEr} will receive alerts from the broker system and perform source classification. Then, it will send this information to \texttt{SkyPortal} and suggest a follow-up spectroscopy using SEDM. The resulting spectra will be fed back into \texttt{AppleCiDEr} for additional inference and refinement.

 We also plan to extend this capability to LSST transients in the near future, specifically within \texttt{BABAMUL}, the public version of \texttt{BOOM} processing LSST alerts. However, generalization to LSST requires careful domain adaptation, particularly to accommodate differences in filter systems between ZTF and LSST, as highlighted by \citet{Muthukrishna2019}.

These enhancements aim to solidify \texttt{AppleCiDEr} not just as a classifier, but as a flexible and extensible engine for discovery in the time-domain sky. \texttt{AppleCiDEr} ultimately provides a scalable and operational framework that bridges the divide between research-grade classifiers and real-time survey pipelines. Validated on more than 110,000 real alerts, it demonstrates the power of unified multimodal learning for immediate classification and retrospective discovery. With its public release, we invite the broader astronomical community to build on this work and shape the next generation of multimessenger science.

\section*{Acknowledgments}
The UMN authors acknowledge support from the National Science Foundation with grant numbers PHY-2117997, PHY-2308862 and PHY-2409481.

N.R. is supported by DoE award \#DE-SC0025599. Zwicky Transient Facility access for N.R. was supported by Northwestern University and the Center for Interdisciplinary Exploration and Research in Astrophysics (CIERA).

Based on observations obtained with the Samuel Oschin Telescope 48-inch and the 60-inch Telescope at the Palomar Observatory as part of the Zwicky Transient Facility project. ZTF is supported by the National Science Foundation under Grants No. AST-1440341, AST-2034437, and currently Award \#2407588. ZTF receives additional funding from the ZTF partnership. Current members include Caltech, USA; Caltech/IPAC, USA; University of Maryland, USA; University of California, Berkeley, USA; University of Wisconsin at Milwaukee, USA; Cornell University, USA; Drexel University, USA; University of North Carolina at Chapel Hill, USA; Institute of Science and Technology, Austria; National Central University, Taiwan, and OKC, University of Stockholm, Sweden. Operations are conducted by Caltech's Optical Observatory (COO), Caltech/IPAC, and the University of Washington at Seattle, USA.

SED Machine is based upon work supported by the National Science Foundation under Grant No. 1106171

The Gordon and Betty Moore Foundation, through both the Data-Driven Investigator Program and a dedicated grant, provided critical funding for SkyPortal.

\section*{Appendix A: Vision transformers for spectra classification}
In this section, we present an alternative approach for spectral classification that leverages CNNs on image-based representations of spectra. Motivated by advances in vision transformer models \citep{sharma2025ccsnscoremultiinputdeeplearning, vision_transformers1, vision_transformers2}, we developed a pipeline that renders one-dimensional flux–wavelength spectra as multi-channel images and processes them using a ConvNeXt-based architecture \citep{liu2022convnet2020s}.

Our network is designed to handle multiband time-series data (e.g., $g$, $r$, and $i$ bands) by converting each observation into an image-like format. These image representations are then passed through a ConvNeXt backbone adapted to accept three-channel input, allowing the model to capture both intra-band temporal features and cross-band correlations. The backbone is followed by a custom classification head consisting of fully connected layers, dropout regularization, and a softmax layer for multi-class prediction. Transfer learning is employed by initializing the ConvNeXt weights with pretrained ImageNet parameters \citep{imagenet}, followed by fine-tuning on our astrophysical dataset.

To generate the input images, we transform each one-dimensional flux–wavelength pair into a two-dimensional line plot using \texttt{matplotlib}. We avoid interpolation and resampling in order to retain the native spectral resolution and sampling characteristics of each instrument. All plot elements such as axes, tick marks, and grid lines were removed to eliminate visual bias, leaving only the flux curve on a white background.

To construct a richer, multi-channel input, we apply LOESS smoothing \citep{JACOBY2000577} to each spectrum with different fractional span values. Specifically, we generate five separate plots: the original raw spectrum (unsmoothed), and four smoothed versions with span values of 0.01, 0.05, 0.1, and 0.2. Each curve is min–max normalized and rendered independently. These images are then stacked to form a $(3, H, W)$ tensor compatible with CNN-based architectures.

We experimented with multiple image resolutions ($224{\times}224$, $288{\times}288$, and $384{\times}384$ pixels) and line widths (0.2, 1.0, and 2.0). The best classification performance was obtained using $288{\times}288$ resolution and a line width of 1.0. Moreover, we found that using only a subset of the smoothing channels—specifically, the raw curve along with LOESS-0.05 and LOESS-0.1—provided improved accuracy compared to using all five.

\begin{figure}
    \centering
    \includegraphics[scale=0.40]{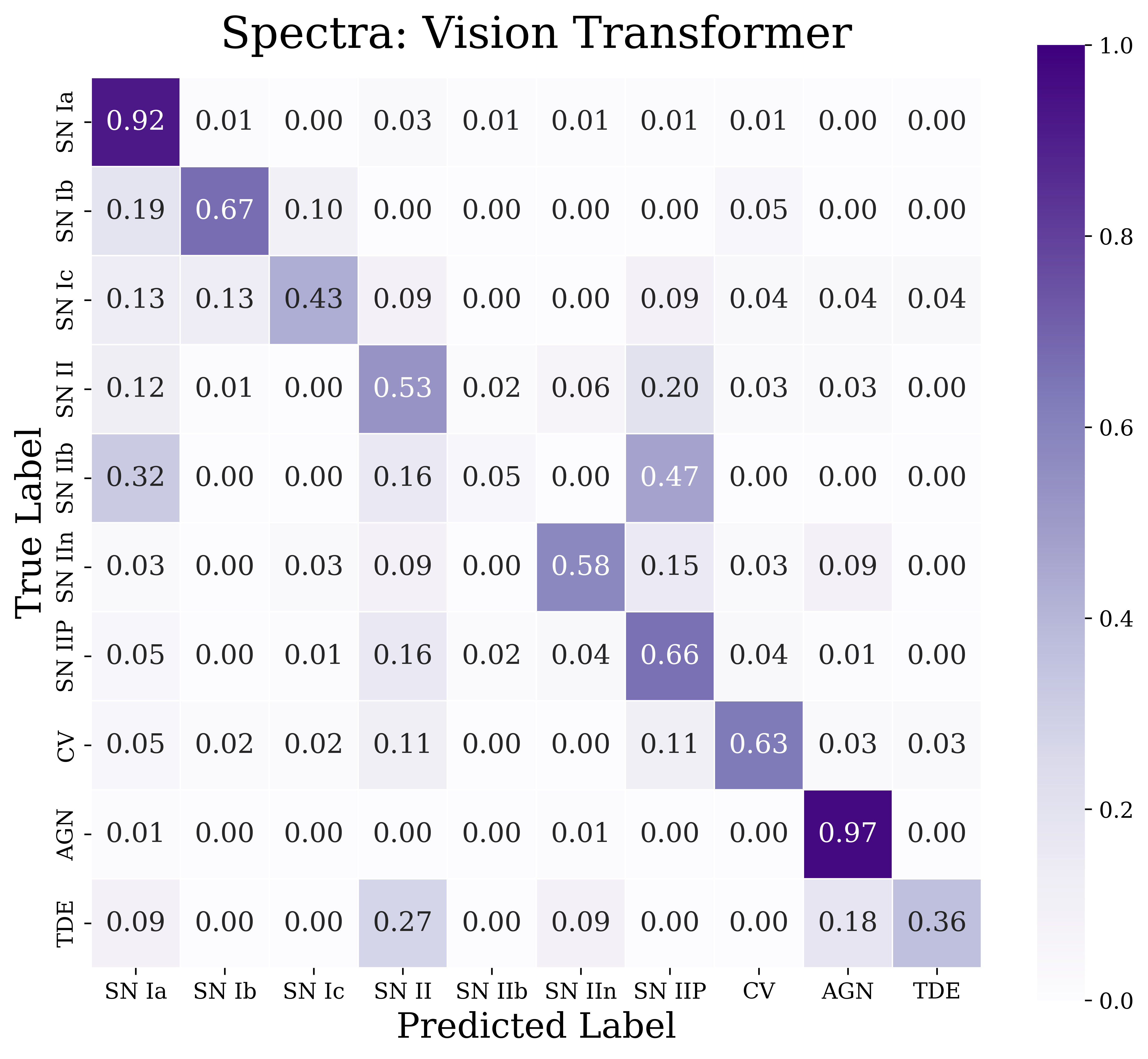}
    \caption{Confusion Matrix using vision transformers.}
    \label{fig:confusion_matrix_spectra_image}
\end{figure}

All rendered images are saved as \texttt{NumPy} arrays and are used directly as input to the classification model. The resulting confusion matrix is shown in Fig.~\ref{fig:confusion_matrix_spectra_image}. Although this approach yields reasonable performance, it exhibits notable confusion between certain classes, and the overall structure of the matrix is less diagonal compared to our final model presented in Sec.~\ref{sec:photometry}.

\end{document}